\documentclass[format=acmsmall, review=false]{acmart}
\usepackage{acm-ec-20}

\usepackage{listings} 
\usepackage{xcolor}

\usepackage{geometry}

\usepackage{amsmath, bm}

\title{Success Stories from a Democratized Experimentation Platform}

\author{Eskil Forsell, Julie Beckley, Simon Ejdemyr, Veronica Hannan, Andy Rhines, Martin Tingley, Matthew Wardrop, Jeffrey Wong (Netflix)
}

\begin{document}

\maketitle


\section{Introduction}
The Netflix experimentation platform (XP) embodies two key principles: (1) that democratizing contributions will allow the platform to scale more efficiently, and (2) that non-distributed computation of causal models will lead to rapid innovations in statistical methodologies \citep{diamantopoulos2020engineering}. As such, we have invested substantial effort in developing a robust experimentation platform that allows for independent contribution of new metrics, reports and statistical methodologies. Innovation on this platform takes place in local R/Python notebooks using standard libraries familiar to data scientist and/or engineers, and can be iterated into production without significant rewrites. The result is a sustainable engineering environment that is inclusive of cutting-edge research.

Over the past two years, the development of this platform has facilitated the diversification of experimentation methodologies at Netflix in response to increasingly nuanced business needs. We started out supporting only t-tests and now have a rich modular modeling framework with implementations including bootstrapping-, linear-, and Bayesian- models.

In the next section we will demonstrate the value of this platform by sharing four recent additions to the experimentation ecosystem that leveraged the democratized contribution model. Although each addition was primarily focused on solving a specific set of real-world business problems, each was readily generalizable and consequently enriched the entire ecosystem, allowing for innovations across business-verticals. We then conclude with a discussion of key points,  and highlight how this framework reduces the cost of exploring novel methodologies developed in tandem with the academe.

\section{Business problems and their solutions}

We describe four challenges faced by the business that led us to implement new methods for analyzing experiments. Although the platform supports any statistical model for estimating treatment effects, the newly contributed models all build on some variation of a linear model, specified as:
\begin{equation}\label{eq:baseOLS}
    y_i=\mathbf{m}_i^\top\cdot\bm{\beta}+\varepsilon_i
\end{equation}
where $y_i$ is the outcome of interest, $\mathbf{m}_i$ is member $i$’s feature vector and $\varepsilon_i$ is an error term. Linear models are flexible and can represent a versatile set of statistical models \citep{lindelov2019}. For example, when $\mathbf{m}_i$ includes an intercept and an indicator for a binary treatment, the model is equivalent to a t-test \citep{wong2019efficient}. To estimate these models we employ the computational strategy outlined in \citep{wong2019efficient} which leverages efficient numerical computing, data compression to decrease data volume, and fast linear algebra, often making it orders of magnitude faster than standard implementations.

\subsection{Adjusting for usage bias in quality metrics: Weighted least squares}
Engineering Quality of Experience (QoE) tests measure changes in playback and app performance. The metrics in these experiments include rates, which are frequently highly skewed and may be zero-inflated. Engineering tests also aggregate data to the account grain, but they still want an estimator that accounts for different levels of user engagement. This combination of properties made them unreliable to model with standard OLS regression, and by extension prevented this class of experiments from leveraging most of the features the XP enables for those models, such as covariate adjustment or heterogeneous treatment effects. Instead, these experiments relied on separate bootstrapping methods, and any innovation in this space required a distinct solution and longer development cycles.

To address these problems we made two changes to the XP that unlocked regression for this type of experiment. First, we added support for analytical weights and used Weighted Least Squares (WLS) to estimate our treatment effects. WLS is a minor update to (\ref{eq:baseOLS}) that combines our existing matrix methods with a new objective function that minimizes the weighted residual sum of squares: 
\begin{equation}\label{eq:wss}
    \text{WSS}(\bm{\beta},\bm{\omega}) = \sum_{i=1}^n \omega_i (y_i -\mathbf{m}_i^\top\cdot\bm{\beta})^2.
\end{equation}

WLS works well in this context because our experiment data is aggregated at the account grain. Without weights, the treatment effect we’d measure would not be representative of the subsequent launch of an intervention. Thus, we adjust for different viewing patterns by weighting the account-aggregated value by the metric’s denominator.

To further improve the accuracy of our standard errors, which exhibit heteroskedasticity, we developed a compression-friendly implementation of White standard errors \citep{white1980robust}.

This combination of changes provided us with an unbiased estimator with reliable coverage, and made regression (as well as all regression-based features) accessible to a new class of metrics.

These changes had an immediate impact on the quality of insights we derived from our experiments. WLS and White standard errors were able to leverage the computational advances that already existed on the platform, making these new methods easy to integrate with our existing infrastructure and highly performant. Up until these changes landed, we were only able to compute the average treatment effect for several core metrics. Now, we can use covariate adjustment to increase our precision, and HTE for a more nuanced understanding of our intervention across several dimensions of interest.

\subsection{Targeted treatments require targeted analysis: Quantile Estimation}
In the streaming domain, an engineering team might change the adaptive streaming engine \citep{ekanadham2018mlstreaming} to accelerate video startup for members experiencing larger delays. This problem is more general across the different business verticals at Netflix, as treatments are often designed with the intent to improve certain quantiles of a metric distribution. In these cases we would not be able to detect movements in the tail with a simple average treatment effect model. We prototyped a novel quantile bootstrapping solution in order to provide inference of quantile treatment effects \citep{beckley2019data}. This solution allowed decision makers to understand whether they moved specific percentiles of metric distributions, generating insights not possible with inference solely on the mean. By using our core concepts -- compression, sparse matrix algebra implemented in C++, and casting methods into a regression-like setting -- we were able to implement our work in the experimentation platforms’s causal model framework. This enabled the computation to be done sufficiently fast and at scale, for all experiments, each with a large number of metrics. 

We can also cast the quantile inference problem more directly in a regression framework \citep{koenker2001quantile} to make use of additional shared features of the platform such as covariate adjustment and conditional effect estimation. Under the hood, estimation for the $\tau$th quantile is done by minimizing
\begin{equation}\label{eq:qr}
   \min_{\beta} \sum_{i=1}^N\rho_\tau\left(y_i - \mathbf{m}_i^\top\cdot\bm{\beta} \right ),
\end{equation}
where $\rho_\tau$ is the tilted absolute loss,
\begin{equation}
\rho_{\tau}(y)=y\left(\tau-\mathbb{I}_{(y<0)}\right).
\end{equation}
Despite the vast algorithmic differences, for the data scientist using quantile regression involves only a minor change from the familiar OLS regression framework. Building both paths has enabled us to perform rapid inference across many metrics when necessary via the bootstrap, and when needed to obtain more precise treatment effect estimates or model nuances of the data via quantile regression.

\subsection{Identifying winning treatments faster and with more precision: Bayesian shrinkage}
Bayesian estimators aim to address at least two challenges encountered in various applications at Netflix. First, selecting winning treatment arms based on unregularized treatment effects is well-known to induce magnitude errors \citep{gelman2014beyond}. Such errors may lead decision makers — who have to weigh the cost and benefits of shipping a new product experience — to overestimate the benefits of an intervention. Second, early stopping of experiments once a winning treatment has been identified facilitates faster experimentation. However, it can be difficult to control false discovery in an early stopping regime while also promoting true discovery. Bayesian estimators can address each challenge by shrinking treatment effects to more plausible values (reducing magnitude errors) and by providing more actionable and stable summaries of uncertainty (facilitating early stopping without adjustments for peeking).

The modular structure of the Netflix XP made it possible to leverage the highly-optimized regression methods that were already on the platform to implement a rich Bayesian model, with support for multiple correlated outcomes, multiple treatment arms, and covariate-adjustment. In particular, the estimator can take advantage of any model described by equation (1), such as t-test and covariate-adjusted OLS, to estimate the likelihood component of the Bayesian model. It then uses informative priors estimated from past experiments to shrink the likelihood estimates and produce various posterior quantities of interest, such as the probability that a treatment arm is the best arm in the experiment. Backtesting the approach on past experiments, we have found reductions to magnitude errors; more stable (relative to p-values) assessments of uncertainty; and overall precision improvements that in some cases are commensurate with including highly-predictive pre-treatment covariates in the model.

\subsection{Understanding dynamics: Treatment effects across time}
Insights into how treatment effects evolve over time help us build intuition into the mechanisms behind average differences and make well informed decisions about which experiences members are most likely to appreciate. To generate these insights our models need to estimate different types of shapes in the treatment effect trend, for example trending positively over time, diminishing over time, or flat over time. To standardize and improve how this is done we consider the special case (with some abuse of notation):
\begin{equation}\label{eq:dte}
    y_{i,t}=\alpha
        +\mathbf{w}_i\bm{\beta}_1
        +\mathbf{x}_i\bm{\beta}_2
        +\mathbf{f}(t)\bm{\beta}_3
        +\mathbf{w}_i\mathbf{x}_i\bm{\beta}_4
        +\mathbf{w}_i\mathbf{f}(t)\bm{\beta}_5
        +\mathbf{x}_i\mathbf{f}(t)\bm{\beta}_6
        +\mathbf{w}_i\mathbf{x}_i\mathbf{f}(t)\bm{\beta}_7
        + \varepsilon_{i,t}
\end{equation}
where $\mathbf{w}_i$ is a treatment indicator, $\mathbf{f}(t)$ a time variable, and $\mathbf{x}_i$ a vector of covariates. Using this single model, we can combine coefficients to estimate (conditional) average, cumulative and daily differences from a single model for many types of metrics. These effects provide a full suite of statistics for business intelligence that can analyze segments and trends.

Because we observe the same member across multiple time periods in this setting the data points are not necessarily i.i.d. To account for this when estimating the model we use cluster-robust covariances \citep{cameron2010robust} that capture the within-member correlation in errors. By leveraging the solution in \citep{wong2020yoco} we achieve compute times on the order of minutes even on panel data for millions of members. Estimating a single model in this way allows us to directly test differences between daily effects with minimal assumptions about the correlation structure. This is useful as it helps us assess whether perceived trends are likely to represent true trends or not.

Finally, the unified model specification in (\ref{eq:dte}) allows users to access treatment effects for different member segments across different periods of time within seconds once the model has been estimated. This makes it easy to go beyond average treatment effects to explore novelty effects, or even the impact of an external event such as a holiday, or recently, a global pandemic.

\section{Conclusion}
This paper highlights the success of our strategy of democratization and efficient computation. This is exemplified by the organic addition of new models that: solve tangible business problems, were contributed by experts from different business areas, and are regularly used in production.

One particularly exciting aspect of our democratized contribution workflows is how often there are synergies between different business verticals, and thus how often verticals benefit passively from contributions made by other verticals. For example, the work on weighted least squares required by one vertical simultaneously allowed for inverse propensity scoring in bandit algorithms used in other verticals.

We are also excited about the prospect of sharing and exploring novel methodologies with the academic community. Our modular and non-distributed computational framework allows us to easily plug in arbitrary algorithms written in Python, R or C/C++, and we are always interested in new approaches that may better inform business strategies. We look forward to continuing to share our own experiences applying these methodologies to real-world data and business decision-making, and to learn from the experiences of our colleagues in both industry and academe.

\bibliographystyle{ACM-Reference-Format}
\bibliography{references}

\end{document}